# MIMO Downlink Scheduling with Non-Perfect Channel State Knowledge

Hooman Shirani-Mehr, Giuseppe Caire and Michael J. Neely

### Abstract

Downlink scheduling schemes are well-known and widely investigated under the assumption that the channel state is perfectly known to the scheduler. In the multiuser MIMO (broadcast) case, downlink scheduling in the presence of non-perfect channel state information (CSI) is only scantly treated. In this paper we provide a general framework that addresses the problem systematically. Also, we illuminate the key role played by the channel state prediction error: our scheme treats in a fundamentally different way users with small channel prediction error ("predictable" users) and users with large channel prediction error ("non-predictable" users), and can be interpreted as a near-optimal *opportunistic* time-sharing strategy between MIMO downlink beamforming to predictable users and space-time coding to non-predictable users. Our results, based on a realistic MIMO channel model used in 3GPP standardization, show that the proposed algorithms can significantly outperform a conventional "mismatched" scheduling scheme that treats the available CSI as if it was perfect.

The authors are with the Ming Hsieh Department of Electrical Engineering, University of Southern California, Los Angeles, CA 90089 USA. E-mail: shiranim@usc.edu, caire@usc.edu, mjneely@usc.edu

# I. INTRODUCTION

Under perfect knowledge of the downlink channels, the resource allocation problem in a Multiuser MIMO (MU-MIMO) downlink has been widely investigated under various precoding and beamfomring schemes (see for example [1], [2], [3], [4], [5], [6], [7], and references therein). In practice, the Channel State Information (CSI) is obtained through some form of training and feedback. In Time-Division Duplexing (TDD) systems, the base station (BS) can learn the downlink channel coefficients in "open-loop" mode, by exploiting the uplink pilot symbols and channel reciprocity (e.g., [8], [9]). In Frequency-Division Duplexing (FDD) systems, since uplink and downlink take place in widely separated frequency bands, the downlink channel coefficients must be learned in "closed loop" mode, via some explicit CSI feedback scheme (e.g., [10], [11] and references therein). In both cases, the CSI available to the BS can be seen as some sort of "noisy" version of the true channel coefficients.

The key impact of CSI quality on the performance of MU-MIMO is evidenced in the relevant regime of medium-to-high SNR. In [10] it is shown that the gap between the sum capacity under perfect CSI and the sum-rate achievable by a simple linear Zero-Forcing Beamforming (ZFBF) scheme with non-perfect CSI takes on the form $\Delta R = \min\{M, K\} \log(1 + \kappa_1 \sigma_e^2 \mathsf{SNR}) + \kappa_2$, where $\kappa_1, \kappa_2$ are constants that depend on the particular CSI training and feedback scheme used, and where $\sigma_e^2$ denotes the Mean-Square Error (MSE) between the true channel coefficients and the CSI available to the BS. Since the sum-rate of a MU-MIMO downlink channel in high-SNR behaves like $\min\{M, K\} \log(1 + \kappa \mathsf{SNR})$, for some constant $\kappa$, it follows that $\sigma_e^2$ must decrease at least as fast as $\mathsf{SNR}^{-1}$ in order to preserve the optimal $O(\min\{M, K\} \log \mathsf{SNR})$ increase of the sum-rate with SNR. Under both TDD and FDD, the main source of CSI estimation error consists of the delay introduced by the estimation/feedback scheme in the presence of time-varying wireless channels [8], [10], [11], [12], [13]: even after neglecting all other sources of suboptimality, such as channel state quantization, feedback errors, and so on, the MSE $\sigma_e^2$ cannot be less than the channel prediction error from noisy pilot symbols. This estimation-theoretic quantity represents a fundamental lower bound to the accuracy of CSI.

We performed an extensive study of MIMO channel prediction based on the 3GPP Spatial Channel Model given in [14]. Our results, summarized in Section V-B, show that channel prediction is generally quite accurate with the exception of a specific class of channels characterized by a large Doppler spread (high user mobility) and clustered angles of arrival. For such channels, the channel prediction MSE is very large, no matter which prediction method is used, as reflected by a Cramer-Rao bound analysis not included in this work for the sake of space limitation [15]. These results suggest that users can be classified according to their channel prediction MSE, and that this classification effectively reduces to only two extreme classes of "non-predictable" (high-mobility and clustered angle of arrivals) and "predictable" (all other cases). This simplification is instrumental to the main contribution of this paper: a simple and efficient MU-MIMO downlink scheduling scheme that takes explicitly into account the CSI quality.

Downlink scheduling aims at making the system operate at a desired point on the *ergodic* (or long-term average) achievable rate region of the system, for a given *physical layer* signaling scheme. The operating point reflects some form of "fairness," corresponding the maximization of a concave non-decreasing utility function of the ergodic rates. Although a direct maximization is typically hopelessly complicated, the optimal point is implicitly achieved using a stochastic optimization approach [16], [17], [18]. We solve this problem for the case of MU-MIMO with non-perfect CSI in Section III. Then, based on the general solution, we find a *practical* simplified scheduling policy under the assumption, motivated before, that the users can be partitioned into two classes with either small or large channel prediction MSE. The resulting scheduling algorithm can be regarded as an opportunistic MIMO "multi-mode" scheme that selects at each scheduling slot either a MU-MIMO downlink beamforming mode that performs spatial multiplexing to a subset of predictable users, or a single-user space-time coding mode that serves a single selected non-predictable user.

With respect to existing literature, we notice that downlink scheduling with non-perfect CSI has been treated mainly in the case where all users have the same CSI quality. *Static* mode-switching criteria have been studied for example in [19], [13] where the number of users to be simultaneously served is optimized depending on the CSI quality and channel SNR. In contrast, the present work presents a *dynamic* scheduling policy that can handle users with very different CSI qualities at the same time, and allocates opportunistically the signaling modes (namely: spatial multiplexing and space-time coding) over time and across the users. The fundamental role of channel prediction in downlink scheduling schemes was noticed before, e.g., in [13], [12]. In particular, [12] proposes a channel-predictive proportional fair scheduling rule, without analytical proof, for the scalar (not MIMO) case. In comparison with these works, here we provide a general framework for downlink scheduling with non-perfect CSI that applies to MU-MIMO and to a wide class of fairness utility functions. Also, we present novel rigorous results on system stability and performance bounds of the proposed scheduling algorithms.

Numerical results are provided for two relevant fairness utility functions reflecting proportional fairness and max-min fairness (referred to as "hard-fairness"). It should be noticed, though, that our framework can be applied to any concave non-decreasing utility function. Results based on a realistic channel model [14] and actual channel state prediction algorithms (see details in Section V) show that the proposed approach achieves very significant improvement with respect to a conventional mismatched scheme that treats the available CSI as if it was perfect.

## II. SYSTEM SET-UP

We consider a MIMO downlink channel with a BS equipped with $M$ antennas and $K$ single-antenna UTs. The channel is assumed frequency flat[1] and constant over "slots" of length $T \gg 1$ symbols (block-fading model). The received complex baseband discrete-time signal at the $k$-th UT during block $t$ is described by

$$y_{k,i}(t) = \mathbf{h}_k^{\mathsf{H}}(t)\mathbf{x}_i(t) + z_{k,i}(t), \quad i = 1, \dots, T \tag{1}$$

where $^{\mathsf{H}}$ denotes *Hermitian transpose*, $t$ tics at the slot rate, $i$ tics at the symbol rate, $k$ denotes the user index, $\mathbf{h}_k(t) \in \mathbb{C}^M$ is the channel vector from the BS antenna array to the $k$-th receiver antenna, $\mathbf{x}_i(t) \in \mathbb{C}^M$ is the transmit signal vector transmitted at symbol interval $i$ of slot $t$, and $z_{k,i}(t) \sim \mathcal{CN}(0, N_0)$ is the corresponding additive white Gaussian noise (AWGN). We collect all channel vectors into a channel state matrix $\mathbf{H}(t) = [\mathbf{h}_1(t), ..., \mathbf{h}_K(t)] \in \mathbb{C}^{M \times K}$. Without loss of fundamental generality, we assume that the channel coefficients have mean 0 and variance 1 (e.g., in the case of Rayleigh fading). At the beginning of each slot $t$, the BS has knowledge of the CSI $\widehat{\mathbf{H}}(t) = [\widehat{\mathbf{h}}_1(t), ..., \widehat{\mathbf{h}}_K(t)] \in \mathbb{C}^{M \times K}$, obtained by some form of training, channel prediction and feedback, as discussed in Section I. We assume that $\mathbf{H}(t)$ and $\widehat{\mathbf{H}}(t)$ are jointly stationary and ergodic matrix-valued processes. For convenience, we also assume that $\widehat{\mathbf{H}}(t)$ is a sufficient statistic for the *causal* estimation of $\mathbf{H}(t)$ from the CSI process $\{\widehat{\mathbf{H}}(t)\}$.

While the capacity region of the MIMO-BC in the *perfect* CSI case (i.e., for $\widehat{\mathbf{H}}(t) = \mathbf{H}(t)$) is well-known [20], the case of imperfect CSI is still open although outer bounds and achievability lower bounds exist. In this work we focus on a simple physical layer signaling scheme based on linear precoding and independently generated Gaussian user codes. Nevertheless, the general scheduling framework developed in this work can be easily extended to other MU-MIMO downlink schemes, such as Tomlinson-Harashima precoding [3], Vector Precoding [21] and Dirty-Paper Coding [20], [22].

---

[1]The generalization to MIMO-OFDM and frequency selective fading is immediate.

With linear precoding, each $k$-th user codeword is a $M \times T$ space-time array denoted by $\mathbf{U}_k(t) = \{\mathbf{u}_{k,i}(t) : i = 1, \ldots, T\}$. The signal vector transmitted at symbol interval $i$ of slot $t$ is given by $\mathbf{x}_i(t) = \sum_{k=1}^{K} \mathbf{u}_{k,i}(t)$. In the following, we let $(\mathbf{H}, \widehat{\mathbf{H}})$ denote a pair of jointly distributed random matrices with the same joint distribution of $(\mathbf{H}(t), \widehat{\mathbf{H}}(t))$ (independent of $t$ by stationarity). A linear precoding *signaling scheme* is defined as a possibly randomized function $\gamma$ such that

$$\gamma(\widehat{\mathbf{H}}) \triangleq (\mathbf{\Sigma}_1(\widehat{\mathbf{H}}), \ldots, \mathbf{\Sigma}_K(\widehat{\mathbf{H}}), \mathbf{r}(\widehat{\mathbf{H}}))$$

where $\mathbf{\Sigma}_k(\widehat{\mathbf{H}})$ is the spatial-domain transmit covariance matrix of user $k$, and $\mathbf{r}(\widehat{\mathbf{H}})$ is a transmit rate allocation vector. Then, upon observation of the CSI $\widehat{\mathbf{H}}(t)$, the signaling scheme $\gamma$ chooses for each user $k$ a Gaussian generated codebook of rate $r_k(\widehat{\mathbf{H}}(t))$, where the codewords $\mathbf{U}_k(t)$ have i.i.d. columns generated according to the Gaussian distribution $\mathcal{CN}(\mathbf{0}, \mathbf{\Sigma}_k(\widehat{\mathbf{H}}(t)))$. We say that a scheme $\gamma$ is feasible with respect to the power constraint $P$ if $\sum_{k=1}^{K} \mathrm{tr}\left(\mathbf{\Sigma}_k(\widehat{\mathbf{H}})\right) \leq P$ with probability 1. The set of all feasible schemes is denoted by $\Gamma(P)$. For a given $\gamma$ and CSI value $\widehat{\mathbf{H}}$, the set of users $k$ such that $\mathrm{tr}(\mathbf{\Sigma}_k(\widehat{\mathbf{H}})) > 0$ is called the *active set*, and will be denoted by $\mathcal{U}_\gamma(\widehat{\mathbf{H}})$.

The above definition of $\gamma$ encompasses in full generality *all* linear precoding strategies based on Gaussian random coding, ranging from beamforming to space-time coding. For later use, we recall here two well-known choices for the transmit covariance matrices that will be essential in the practical scheduling policy of Section IV:

1) A popular choice for MU-MIMO linear precoding consists of computing ZFBF "steering vectors" by treating the CSI $\widehat{\mathbf{H}}$ as if it was the true channel matrix (see for example [8], [10], [23] and references therein). In our notation, this corresponds to choosing an active set $\mathcal{U}_\gamma(\widehat{\mathbf{H}})$ of size not larger than $\mathrm{rank}(\widehat{\mathbf{H}})$ and, rank-1 transmit covariance matrices $\mathbf{\Sigma}_k(\widehat{\mathbf{H}}) = p_k \mathbf{v}_k \mathbf{v}_k^{\mathsf{H}}$ where $p_k > 0$ is the transmit power allocated to user $k$, and $\mathbf{v}_k$ is a unit-length vector obtained by calculating the Moore-Penrose pseudo-inverse

$$\widehat{\mathbf{H}}^\dagger(\mathcal{U}_\gamma) = \widehat{\mathbf{H}}(\mathcal{U}_\gamma) \left( \widehat{\mathbf{H}}^\mathsf{H}(\mathcal{U}_\gamma) \widehat{\mathbf{H}}(\mathcal{U}_\gamma) \right)^{-1} \tag{2}$$

of the matrix $\widehat{\mathbf{H}}(\mathcal{U}_\gamma)$ with columns $\{\widehat{\mathbf{h}}_j : j \in \mathcal{U}_\gamma\}$, and taking the normalized column of $\widehat{\mathbf{H}}^\dagger(\mathcal{U}_\gamma)$ corresponding to user $k$. In particular, $\mathbf{v}_k$ is orthogonal to all $\{\widehat{\mathbf{h}}_j : j \in \mathcal{U}_\gamma(\widehat{\mathbf{H}}), j \neq k\}$.

2) At the other extreme of the range of possible linear precoding signaling schemes we find the classical space-time coding to a single user [13], [24], [25]. In our notation, this corresponds to choosing an active set $\mathcal{U}_\gamma(\widehat{\mathbf{H}})$ of size 1 and the transmit covariance matrix $\boldsymbol{\Sigma}_k(\widehat{\mathbf{H}}) = (P/M)\mathbf{I}$ for the only $k \in \mathcal{U}_\gamma(\widehat{\mathbf{H}})$. Interestingly, ZFBF serves simultaneously up to $M$ active users, each with a rank-1 transmit covariances matrix, while space-time coding serves just one active user, with a rank-$M$ transmit covariance matrix.

For a fixed set of transmit covariance matrices $(\boldsymbol{\Sigma}_1, \ldots, \boldsymbol{\Sigma}_K)$, a linear precoding scheme yields a Signal-to-Interference plus Noise Ratio (SINR) at receiver $k$ given by

$$\mathsf{SINR}_k(\mathbf{H}, \boldsymbol{\Sigma}_1, \ldots, \boldsymbol{\Sigma}_K) \triangleq \frac{\mathbf{h}_k^\mathsf{H} \boldsymbol{\Sigma}_k \mathbf{h}_k}{N_0 + \sum_{j \neq k} \mathbf{h}_k^\mathsf{H} \boldsymbol{\Sigma}_j \mathbf{h}_j} \tag{3}$$

We let $R_k(t)$ denote the effective rate of user $k$ on slot $t$. In general $R_k(t)$ is different from the allocated rate $r_k(\widehat{\mathbf{H}}(t))$ since CSI is not perfect. As far as *rate allocation* is concerned, we consider the following two cases:

*1) Outage rates:* following standard information theoretic arguments (see [26] and references therein), under slot-by-slot coding and decoding, receiver $k$ can reliably decode a rate $r_k$ provided that no *information-outage* occurs, i.e., provided that $r_k$ is smaller than the mutual information $I_k(\mathbf{H}, \boldsymbol{\Sigma}_1, \ldots, \boldsymbol{\Sigma}_K) \triangleq \log\left(1 + \mathsf{SINR}_k\left(\mathbf{H}, \boldsymbol{\Sigma}_1, \ldots, \boldsymbol{\Sigma}_K\right)\right)$. As a consequence, for a given signaling scheme $\gamma$ we define the outage rate as the random variable:

$$R_k(\mathbf{H}, \gamma(\widehat{\mathbf{H}})) = r_k(\widehat{\mathbf{H}}) \times 1\left\{ r_k(\widehat{\mathbf{H}}) < I_k(\mathbf{H}, \boldsymbol{\Sigma}_1(\widehat{\mathbf{H}}), \ldots, \boldsymbol{\Sigma}_K(\widehat{\mathbf{H}})) \right\} \tag{4}$$

where $1\{\mathcal{A}\}$ is the indicator function of an event $\mathcal{A}$.

*2) Optimistic rates:* in this case, we *assume* that some genie-aided rate adaptation scheme is able to achieve an effective instantaneous rate equal to the mutual information:

$$R_k(\mathbf{H}, \gamma(\widehat{\mathbf{H}})) = I_k(\mathbf{H}, \boldsymbol{\Sigma}_1(\widehat{\mathbf{H}}), \ldots, \boldsymbol{\Sigma}_K(\widehat{\mathbf{H}})) \tag{5}$$

The system model underlying the outage rate assumption consists of standard ARQ protocol that removes $R_k(t) = r_k(\widehat{\mathbf{H}}(t))$ bits/channel use from the transmission buffer of active user $k$ if no outage occurs. The system model underlying the optimistic rate assumption corresponds to an idealized fast rate adaptation scheme (see for example [27], [28]). Any practical rate adaptation scheme yields performance in between the outage and the optimistic rates defined above. Therefore, these two extreme cases are relevant in the sense that they provide upper and lower bounds to practical adaptive rate schemes. Under either one of the above assumptions, we let the effective rate be $R_k(t) = R_k(\mathbf{H}(t), \gamma(\widehat{\mathbf{H}}(t)))$, given by (4) or by (5).

## III. Optimal downlink scheduling

The achievable ergodic rate region $\mathcal{R}$, for a given set of feasible physical layer signaling schemes, is defined as the closure of the convex hull of all achievable ergodic rate points. Under a fixed signaling scheme $\gamma \in \Gamma(P)$, user $k$ is served with a long-term average rate $\bar{R}_k = \lim_{t\to\infty} \frac{1}{t} \sum_{\tau=0}^{t-1} R_k(\mathbf{H}(\tau), \gamma(\widehat{\mathbf{H}}(\tau))) = \mathbb{E}[R_k(\mathbf{H}, \gamma(\widehat{\mathbf{H}}))]$, where convergence is with probability 1 because of ergodicity. Since time-sharing between any set of feasible signaling schemes is also a feasible scheme, we have:

$$\mathcal{R} = \text{coh} \bigcup_{\gamma \in \Gamma(P)} \left\{ \bar{\mathbf{R}} \in \mathbb{R}_+^K : \bar{R}_k \leq \mathbb{E}\left[ R_k(\mathbf{H}, \gamma(\widehat{\mathbf{H}})) \right], \ \forall \ k \right\} \tag{6}$$

where "coh" denotes "closure of the convex hull" and where the expectation is with respect to the joint probability distribution of $(\mathbf{H}, \widehat{\mathbf{H}})$ and $\gamma$ (for randomized signaling schemes).

We consider an "infinite backlog" situation where all the data to be transmitted are available at the BS. The goal of the downlink scheduler is to maximize some concave entrywise non-decreasing utility function $g(\cdot)$ of the user individual ergodic rates $\bar{\mathbf{R}} = (\bar{R}_1, \ldots, \bar{R}_K)$.[2] The problem that we wish to solve is:

$$\text{maximize} \ \ g(\bar{\mathbf{R}}), \qquad \text{subject to} \ \ \bar{\mathbf{R}} \in \mathcal{R} \tag{7}$$

Suppose that the solution $\bar{\mathbf{R}}^\star$ of (7) is found. Then, by definition, there exists a (possibly randomized) signaling strategy that achieves $\bar{\mathbf{R}}^\star$. A *feasible scheduling policy* is an algorithm that chooses at each time $t$ some physical layer signaling scheme $\gamma \in \Gamma(P)$, based on the history of all past transmissions and arrivals, on the observation of the CSI and on the knowledge of the joint statistics of all variables in the system. We are interested in finding an explicit scheduling policy (denoted for brevity by $\gamma^\star$) that achieves $\bar{\mathbf{R}}^\star$. Despite the fact that (7) is a convex optimization problem, a direct solution is generally overly complicated since $\mathcal{R}$ does not admit in general a simple characterization. For example, $\mathcal{R}$ is generally not a polytope, and may be described by an uncountable number of linear constraints (supporting hyperplanes).

---

[2]By entrywise non-decreasing we mean that for all $\mathbf{r} \in \mathbb{R}_+^K$ and $\boldsymbol{\delta} \in \mathbb{R}_+^K$, $g(\mathbf{r}) \geq g(\mathbf{r} + \boldsymbol{\delta})$. Also, recall that a concave function is continuous in the interior of its domain. Without loss of generality, we consider $g(\cdot)$ with domain $\mathbb{R}_+^K$.

Fortunately, we can use the framework of [16] and obtain a dynamic scheduling policy that operates arbitrarily closely to the optimal point $\bar{\mathbf{R}}^\star$. This is obtained in two steps: first, a dynamic scheduling policy that achieves the stability of transmission queues whenever the arrival rates are inside $\mathcal{R}$ is obtained. Then, we build "virtual queues" driven by appropriate "virtual arrival processes," such that their arrival rates are as close as desired to the desired rate point $\bar{\mathbf{R}}^\star$. Our analysis extends the results in [16] to this new context and also provides a new and tighter bounding analysis for the queues, particularly for general (possibly negative) concave utilities that include the proportional fairness utility. We note that it may be possible to pursue utility optimization using the alternative stochastic approximation and fluid transformation approaches in [29], [30], [31], [32], although these approaches may not yield explicit queue bounds. Further, the stochastic approximation techniques in [29], [30] use an infinite running time average of transmission rates, whereas our approach does not require an infinite running time average and can thus adapt to system changes.

### A. System stability

Suppose that the data to be transmitted to user $1, \ldots, K$ arrive to the BS according to a stationary and ergodic vector-valued process $\mathbf{A}(t) = (A_1(t), \ldots, A_K(t))$, with rate vector $\boldsymbol{\lambda} = \mathbb{E}[\mathbf{A}(t)]$ (expressed in bit/channel use) and such that $0 \le A_k(t) \le A_{\max}$, $\forall\, t$, for some constant $A_{\max} < \infty$. The BS maintains a transmission queue for each user, and we let $Q_k(t)$ denote the size of the $k$-th queue buffer at the beginning of slot $t$. As described in Section II, $R_k(t)$ bit/channel use are removed from queue $k$ during slot $t$, i.e., $R_k(t)$ represents the instantaneous "service rate" of the $k$-th queue. Defining $\mathbf{Q}(t) = (Q_1(t), \ldots, Q_K(t))$ and $\mathbf{R}(t) = (R_1(t), \ldots, R_K(t))$, the queues evolution is described by the stochastic difference equation[3]

$$\mathbf{Q}(t+1) = \max\{\mathbf{0}, \mathbf{Q}(t) - \mathbf{R}(t)\} + \mathbf{A}(t) \tag{8}$$

We have the following definition [16]:

---

[3]The function $\max\{\cdot, \cdot\}$ is applied componentwise to vectors.

*Definition 1:* A discrete-time queue $Q_k(t)$ is *strongly stable* if $\limsup_{t\to\infty} \frac{1}{t} \sum_{\tau=0}^{t-1} \mathbb{E}[Q_k(\tau)] < \infty$. The system is strongly stable if all queues $k = 1, \ldots, K$ are strongly stable. $\Diamond$

For convenience, throughout this paper we use the term "stability" to refer to strong stability. It can be shown [16] that if $Q_k(t)$ is strongly stable and $A_k(t)$ is uniformly bounded by a finite constant $A_{\max}$, as in our case, then $\lim_{t\to\infty} Q_k(t)/t = 0$ with probability 1 and $\lim_{t\to\infty} \mathbb{E}[Q_k(t)]/t = 0$. These properties are referred to as *rate stability* and *mean-rate stability*, respectively. In particular, rate stability implies that the time average rate of bits going into the queue is equal to the time average rate of bits going out of the queue.

The system stability region is the the closure of the convex hull of all arrival rate points $\boldsymbol{\lambda}$ for which there exists a feasible scheduling policy that achieves system stability [16]. The following result yields both the system stability region and the dynamic scheduling policy that stabilizes the system for any arrival rate point inside the region:

*Theorem 1:* Suppose the arrival vector $\mathbf{A}(t)$ is i.i.d. over slots with each entry uniformly bounded by some finite constant $A_{max}$, and that the joint channel state and CSI pair $\{\mathbf{H}(t), \widehat{\mathbf{H}}(t)\}$ is i.i.d. over slots.[4] For the system defined in Section II, the system stability region coincides with the ergodic rate region $\mathcal{R}$ given in (6). Furthermore, for any arrival rate point $\boldsymbol{\lambda}$ in the interior of $\mathcal{R}$, the system is stabilized by the dynamic scheduling policy $\gamma^*$ defined as follows. For $\mathbf{Q} \in \mathbb{R}_+^K$ and $\widehat{\mathbf{H}} \in \mathbb{C}^{M\times K}$, consider the signaling scheme with covariance matrices $(\boldsymbol{\Sigma}_1^*(\widehat{\mathbf{H}}), \ldots, \boldsymbol{\Sigma}_K^*(\widehat{\mathbf{H}}))$ and rate allocation vector $\mathbf{r}^*(\widehat{\mathbf{H}})$ given by the solution of the weighted sum-rate maximization:

$$
\begin{aligned}
\text{maximize} \quad & \textstyle\sum_{k=1}^{K} Q_k\, \mathbb{E}\left[\, R_k(\mathbf{H}, \boldsymbol{\Sigma}_1, \ldots, \boldsymbol{\Sigma}_K, \mathbf{r})|\widehat{\mathbf{H}}\right] \\
\text{subject to} \quad & \textstyle\sum_{k=1}^{K} \mathrm{tr}(\boldsymbol{\Sigma}_k) \leq P, \quad \boldsymbol{\Sigma}_k \geq 0 \ \forall\, k, \quad \mathbf{r} \geq 0
\end{aligned}
\tag{9}
$$

Then, the dynamic scheduling policy $\gamma^*$ chooses at each time $t$ the signaling scheme defined by (9) for the current queue states (i.e., for $\mathbf{Q} = \mathbf{Q}(t)$) and the current CSI (i.e., for $\widehat{\mathbf{H}} = \widehat{\mathbf{H}}(t)$).

*Proof:* See Appendix I ∎

Interestingly, the weighted sum-rate maximization in (9) defining $\gamma^*$ involves the *conditional expected service rates* for given CSI: in the absence of perfect CSI the BS schedules the users on the basis of the MMSE estimation (conditional mean) of their instantaneous service rates. We conclude this section with a note on the optimal rate allocation in the stability policy $\gamma^*$. Under the optimistic rate assumption, $\mathbf{r}$ is irrelevant since (5) does not depend on $\mathbf{r}$. Under the outage rate assumption, using (4) we obtain the optimal rate allocation $\mathbf{r}^*$ for a given set of covariance matrices and CSI as the solution of (see also [13]):

$$r_k^*(\widehat{\mathbf{H}}) = \arg\max_{r \geq 0} \; r \left[ 1 - \mathbb{P}\left( \log\left(1 + \mathsf{SINR}_k(\mathbf{H}, \boldsymbol{\Sigma}_1, \dots, \boldsymbol{\Sigma}_K)\right) \leq r | \widehat{\mathbf{H}} \right) \right] \tag{10}$$

---

[4]This result and the result of Theorem 2 are stated for the i.i.d. case, but they can be extended to jointly ergodic processes subject to some mild technical conditions by following the technique of [18]. We omit this extension for brevity.

### B. System optimization

Going back to the original problem (7) and following [16], we build "virtual queues" with arrival rate $K$-tuple $\boldsymbol{\lambda}$ arbitrarily close to the desired optimal point $\bar{\mathbf{R}}^\star$, although the latter is not known a priori. Then, using the stability policy $\gamma^*$ applied to the virtual queues, the system necessarily operates at a throughput point $\bar{\mathbf{R}} \geq \boldsymbol{\lambda}$ (componentwise domination). From the monotonicity of $g(\cdot)$ we are guaranteed that the system will operate arbitrarily close to the optimal point.

Specifically, we define $\gamma^\star$ as follows: let $V, A_{\max} > 0$ be suitable constants. At each time $t$, let $A_k(t) = a_k$, where $\mathbf{a} = (a_1, \dots, a_K)$ is the solution of

$$\max_{\mathbf{a}: 0 \leq a_k \leq A_{\max}, \, \forall k} \; V g(\mathbf{a}) - \sum_{k=1}^{K} a_k Q_k(t) \tag{11}$$

Then, for given $\mathbf{Q}(t)$ and CSI $\widehat{\mathbf{H}}(t)$ the signaling scheme $\gamma^*$ given in (9) is applied, resulting in the service rates $\mathbf{R}(t)$. Finally, the virtual queues are updated according to (8), with arrivals $\mathbf{A}(t)$ given by (11) and service rates $\mathbf{R}(t)$. The performance of the scheduling policy $\gamma^\star$ is given by the following:

*Theorem 2:* Suppose the joint channel state and CSI pair $\{\mathbf{H}(t), \widehat{\mathbf{H}}(t)\}$ is i.i.d over slots. Consider the scheduling policy $\gamma^\star$ defined above, for given constants $V > 0$ and $A_{\max} > 0$. Assume that $g(\cdot)$ is concave and entry-wise non-decreasing and that there exists at least one point $\mathbf{r} \in \mathcal{R}$ with strictly positive entries such that $g(\mathbf{r}/2) > -\infty$. Then:

(a) The utility associated with the time average transmission rates achieved by $\gamma^\star$ satisfies:

$$\liminf_{t \to \infty} g\left( \frac{1}{t} \sum_{\tau=0}^{t-1} \mathbb{E}[\mathbf{R}(\tau)] \right) \geq g(\bar{\mathbf{R}}^\star(A_{\max})) - C/V \tag{12}$$

where

$$C \triangleq \frac{K\left(A_{\max}^2 + \mathbb{E}[\log^2(1 + |\mathbf{h}_k(t)|^2 P/N_0)]\right)}{2} \tag{13}$$

and where $\bar{\mathbf{R}}^\star(A_{\max})$ denotes the solution of the problem (7) with the additional constraint $0 \leq \bar{R}_k \leq A_{\max}$ for all $k = 1, \ldots, K$.

(b) For any point $\bar{\mathbf{R}} \in \mathcal{R}$ such that $0 \leq \bar{R}_k \leq A_{\max}$ for all $k$, and for any value $\beta \in [0, 1]$ we have:

$$\limsup_{t \to \infty} \frac{1}{t} \sum_{\tau=0}^{t-1} \sum_{k=1}^{K} \bar{R}_k \mathbb{E}[Q_k(\tau)] \leq \frac{C + V[g(\bar{\mathbf{R}}^\star(A_{\max})) - g(\beta\bar{\mathbf{R}})]}{1 - \beta} \tag{14}$$

Thus, all queues $Q_k(t)$ are strongly stable.

*Proof:* See Appendix I. ∎

Theorem 2 implies that if $A_{\max}$ is sufficiently large, such that $A_{\max} \geq \bar{R}_k^\star$ for all $k$, then:

$$\liminf_{t \to \infty} g\left(\frac{1}{t} \sum_{\tau=0}^{t-1} \mathbb{E}[\mathbf{R}(\tau)]\right) \geq g(\bar{\mathbf{R}}^\star) - C/V.$$

Hence, the control parameter $V$ can be chosen as large as desired to make the achieved utility arbitrarily close to the optimal utility $g(\bar{\mathbf{R}}^\star)$ for the problem (7). This comes with a tradeoff in the virtual queue average sizes that, as seen from (14), grow linearly with $V$. The virtual queue sizes represent the difference between the virtual bits admitted into the queues and the actual bits transmitted, and thus affect the time-scales required for the time averages to become close to their limiting values.

## C. Proportional fairness and hard fairness scheduling

We shall focus on two particularly relevant special cases for the system utility function $g(\cdot)$ that reflect useful forms of *fairness*. The *proportional fairness schedulling* (PFS) is defined by the utility function $g(\bar{\mathbf{R}}) = \sum_k \log \bar{R}_k$ [5], [6]. In this case, the solution of (11) is given by:

$$A_k(t) = \min\left\{\frac{V}{Q_k(t)}, A_{\max}\right\} \tag{15}$$

The *hard fairness scheduling* (HFS), uses the utility function $g(\bar{\mathbf{R}}) = \min\{\bar{R}_1, \ldots, \bar{R}_K\}$. Making use of an auxiliary variable $\rho$, (11) can be re-stated as

$$\max_{0 \leq \rho \leq a_k \leq A_{\max}} V\rho - \sum_{k=1}^{K} a_k Q_k(t) \tag{16}$$

Solving first with respect to a for fixed $\rho$ and then solving with respect to $\rho$ we obtain

$$A_k(t) = \begin{cases} A_{\max} & \text{if } V > \sum_{k=1}^{K} Q_k(t) \\ 0 & \text{else} \end{cases} \tag{17}$$

An interesting and new aspect of Theorem 2 is that part (b) allows us to find explicitly the tighter upper bound on the virtual queue sizes. For example, consider the proportional fairness utility, choose a sufficiently large $A_{\max}$ such that $A_{\max} \geq \bar{R}_k^\star$ for all $k$, and consider the vector $\bar{\mathbf{R}} = \bar{\mathbf{R}}^\star$ in (14). Then, we have $g(\beta \bar{\mathbf{R}}^\star) = K \log(\beta) + g(\bar{\mathbf{R}}^\star)$ and the bound in Theorem 2 part (b) becomes $\frac{C - VK \log(\beta^*)}{1 - \beta^*}$, where $\beta^*$ is the unique solution in $[0, 1]$ of the equation $\log \beta + \frac{1}{\beta} = 1 + \frac{C}{KV}$.

For the hard fairness utility, choosing again $A_{\max} \geq \bar{R}_k^\star$ for all $k$ and letting $\bar{\mathbf{R}} = \bar{\mathbf{R}}^\star$ in the bound, we have $g(\beta \bar{\mathbf{R}}^\star) = \beta g^\star$, where $g^\star$ is the max-min ergodic per-user rate. Hence, the bound in Theorem 2 part (b) becomes simply $C + V g^\star$.

## IV. Algorithms

In order to implement $\gamma^\star$, a simple and easily computable expression for the conditional average service rates $\mathbb{E}[R_k(\mathbf{H}, \mathbf{\Sigma}_1, \ldots, \mathbf{\Sigma}_K, \mathbf{r}) | \widehat{\mathbf{H}}]$ is needed. Unfortunately, this is generally not available even in the case where the joint statistics of $\mathbf{H}$ and $\widehat{\mathbf{H}}$ is known and easy to characterize (e.g., when $\mathbf{H}$ and $\widehat{\mathbf{H}}$ are jointly Gaussian). Another difficulty in the implementation of $\gamma^\star$ is that the weighted sum rate maximization in (9) is generally a non-convex problem for the linear precoding signaling schemes at hand.

To overcome these difficulties, we introduce some approximations. We start by considering the ZFBF scheme reviewed in Section II. In this case, the maximization in (9) reduces to the maximization over all active sets $\mathcal{U} \subseteq \{1, \ldots, K\}$ such that $|\mathcal{U}| \leq M$ of the weighted sum-rate

$$S(\mathcal{U}, \mathbf{p}, \mathbf{Q}, \widehat{\mathbf{H}}) = \sum_{k \in \mathcal{U}} Q_k \mathbb{E}\left[\log\left(1 + \frac{|\mathbf{h}_k^{\mathsf{H}} \mathbf{v}_k|^2 p_k}{N_0 + \sum_{j \in \mathcal{U}: j \neq k} |\mathbf{h}_k^{\mathsf{H}} \mathbf{v}_j|^2 p_j}\right)\middle| \widehat{\mathbf{H}}\right] \tag{18}$$

where $\mathbf{p} = \{p_k\}$ and $\mathbf{v}_k$ are the ZFBF vectors defined in Section II. Notice that $\{\mathbf{v}_k : k \in \mathcal{U}\}$ and $\mathbf{p}$ are functions of $\widehat{\mathbf{H}}$, even though we omit the explicit dependence for the sake of notation simplicity. Also, it is understood that $p_k = 0$ for all $k \notin \mathcal{U}$.

Motivated by the findings on channel state prediction error mentioned in Section I and illustrated in Section V, we make the working assumption that the users can be partitioned into two classes: the subset $\mathcal{K}_{\mathrm{pr}}$ of users with very small channel state prediction error ("predictable users") and the subset $\mathcal{K}_{\mathrm{npr}}$ of users with very large channel state prediction error ("non-predictable users"). In order to develop some intuition, we assume the CSI model $\mathbf{h}_k = \widehat{\mathbf{h}}_k + \mathbf{e}_k$ where the CSI estimation error $\mathbf{e}_k$ is statistically independent of $\widehat{\mathbf{h}}_k$, with mean zero and variance $\sigma_e^2$ per component. Furthermore, we restrict to the on-off power allocation $p_k = P/|\mathcal{U}| \times 1\{k \in$ $\mathcal{U}\}$, that is known to yield near-optimal sum-rate for the optimal choice of the active user set $\mathcal{U}$ and sufficiently large $P/N_0$. We wish to understand whether a given user should be included in the active set in the maximization of (18). For this purpose, we evaluate the gap between the actual service rate of user $k$ and the service rate user $k$ would achieve if the BS has perfect knowledge of $\mathbf{h}_k$. We will evaluate this gap under both assumptions $k \in \mathcal{K}_{\text{pr}}$ (corresponding to $\sigma_e^2 \approx 0$) and $k \in \mathcal{K}_{\text{npr}}$ (corresponding to $\sigma_e^2 \approx 1$), and eventually conclude that if a user $k \in \mathcal{K}_{\text{npr}}$ is to be served, then no other user should be served in the same slot.

Suppose that a genie provides the true channel vector $\mathbf{h}_k$ to the BS. Then perfect zero-forcing to user $k$ is possible. We will denote by $\widehat{\mathbf{H}}_k^{\text{genie}}$ the genie-aided CSI obtained by replacing $\widehat{\mathbf{h}}_k$ with $\mathbf{h}_k$ in the CSI matrix $\widehat{\mathbf{H}}$. The beamforming vectors computed using the genie-aided CSI are denoted by $\{\mathbf{v}_j^{\text{genie}} : j \in \mathcal{U}\}$, and have the property that $\mathbf{h}_k^{\mathsf{H}} \mathbf{v}_j^{\text{genie}} = 0$ for $j \neq k$, since $\mathbf{h}_k$ is known perfectly. In general, $\mathbf{v}_k^{\text{genie}} \neq \mathbf{v}_k$ unless $|\mathcal{U}| = M$.

The conditional expected service rate under the augmented CSI for user $k$ is given by

$$
\begin{aligned}
R_k(\widehat{\mathbf{H}}_k^{\text{genie}}) &= \mathbb{E}\left[\log\left(1 + \frac{|\mathbf{h}_k^{\mathsf{H}} \mathbf{v}_k^{\text{genie}}|^2 P}{N_0|\mathcal{U}| + \sum_{j \in \mathcal{U}: j \neq k} |\mathbf{h}_k^{\mathsf{H}} \mathbf{v}_j^{\text{genie}}|^2 P}\right) \middle| \widehat{\mathbf{H}}, \mathbf{h}_k\right] \\
&= \log\left(1 + \frac{|\mathbf{h}_k^{\mathsf{H}} \mathbf{v}_k^{\text{genie}}|^2 P}{N_0|\mathcal{U}|}\right),
\end{aligned}
\tag{19}
$$

while in the case of the actual CSI we have

$$
R_k(\widehat{\mathbf{H}}) = \mathbb{E}\left[\log\left(1 + \frac{|\mathbf{h}_k^{\mathsf{H}} \mathbf{v}_k|^2 P}{N_0|\mathcal{U}| + \sum_{j \in \mathcal{U}: j \neq k} |\mathbf{e}_k^{\mathsf{H}} \mathbf{v}_j|^2 P}\right) \middle| \widehat{\mathbf{H}}\right]
\tag{20}
$$

Defining the conditional rate-gap as $\Delta R_k(\widehat{\mathbf{H}}, \mathbf{h}_k) = R_k(\widehat{\mathbf{H}}_k^{\text{genie}}) - R_k(\widehat{\mathbf{H}})$, using (19) and (20), the monotonicity of $\log(\cdot)$ and Jensen's inequality, after simple algebra we obtain the rate-gap upper bound:

$$
\Delta R_k(\widehat{\mathbf{H}}, \mathbf{h}_k) \leq \underbrace{\log\left(1 + \frac{|\mathbf{h}_k^{\mathsf{H}} \mathbf{v}_k^{\text{genie}}|^2 P}{N_0|\mathcal{U}|}\right) - \mathbb{E}\left[\log\left(1 + \frac{|\mathbf{h}_k^{\mathsf{H}} \mathbf{v}_k|^2 P}{N_0|\mathcal{U}|}\right) \middle| \widehat{\mathbf{H}}\right]}_{\Theta_k} + \log\left(1 + \frac{\sigma_e^2(|\mathcal{U}| - 1)P}{N_0|\mathcal{U}|}\right)
$$

By the properties of the Moore-Penrose pseudo-inverse (2), we have that $\mathbb{E}[\Theta_k] \geq 0$ where equality holds exactly when $|\mathcal{U}| = M$ and approximately when $\sigma_e^2 \approx 0$. It follows that if $k \in \mathcal{K}_{\text{pr}}$, then $\Delta R_k(\widehat{\mathbf{H}}, \mathbf{h}_k) \approx 0$ with high probability, i.e., for predictable users the gap between perfect and non-perfect CSI is very small, as we may expect. In contrast, if $k \in \mathcal{K}_{\text{npr}}$, then $\mathbf{h}_k \approx \mathbf{e}_k$, independent of $\widehat{\mathbf{h}}_k$. In this case, $\Delta R_k(\widehat{\mathbf{H}}, \mathbf{h}_k) \approx \Theta_k + \log\left(1 + \frac{\sigma_\varepsilon^2(|\mathcal{U}|-1)P}{N_0|\mathcal{U}|}\right)$ grows on average like $\log(P/N_0)$, unless we let $\mathcal{U} = \{k\}$, i.e., we schedule only user $k$.

The above argument leads to the following conclusions: a near-optimal scheduling policy should select at each slot $t$ either a group of predictable users and serve them using ZFBF spatial multiplexing mode, or a single non-predictable user and serve it using space-time coding, that does not require CSI at the transmitter apart from the rate allocation. Operating along these guidelines, in all cases the rate-gap with respect to perfect CSI is a constant that does not grow with $P/N_0$.

For $\mathcal{U} \subseteq \mathcal{K}_{\mathrm{pr}}$, the objective function in (18) becomes

$$S^{\mathrm{pr}}(\mathcal{U}, \mathbf{p}, \mathbf{Q}, \widehat{\mathbf{H}}) \approx \sum_{k \in \mathcal{U}} Q_k \log\left(1 + \frac{|\mathbf{h}_k^{\mathsf{H}} \mathbf{v}_k|^2 p_k}{N_0}\right) \tag{21}$$

where, for each such subset, the ZFBF vectors $\{\mathbf{v}_k\}$ are obtained as the normalized columns of the Moore-Penrose pseudo-inverse (2). Then, the power allocation vector $\mathbf{p}$ is obtained from the standard waterfilling formula [2], [4]. If the number of predictable users is large, the near-optimal user selection algorithm of [4] can be used to avoid the combinatorial search over all $\mathcal{U} \subseteq \mathcal{K}_{\mathrm{pr}}$. For the non-predictable users, the corresponding objective function is given by

$$S^{\mathrm{npr}}(\{k\}, \mathbf{Q}) = \begin{cases} Q_k \mathbb{E}\left[\log\left(1 + \frac{|\mathbf{h}_k|^2 P}{M N_0}\right)\right] & \text{for optimistic rates} \\ Q_k \max_{r \geq 0}\left\{r\left[1 - \mathbb{P}\left(|\mathbf{h}_k|^2 \leq \frac{2^r - 1}{P/(M N_0)}\right)\right]\right\} & \text{for outage rates} \end{cases} \tag{22}$$

Notice that the rate allocation in the outage rate case is actually very simple: it is sufficient to know the cdf of $|\mathbf{h}_k|^2$, which is either known a priori or it can be learned "on-line" from the channel measurements at each UT.

The the proposed simplified scheduling policy can be summarized as follows: for a desired concave non-decreasing utility function $g(\cdot)$ of the ergodic rates, the virtual arrival processes and the corresponding virtual queues are defined in Section III-B, yielding queue buffers $\mathbf{Q}(t)$ at each scheduling slot $t$. The scheduler computes $S^{\mathrm{pr}}_{\max}(t) = \max_{\mathcal{U} \subseteq \mathcal{K}_{\mathrm{pr}}} S^{\mathrm{pr}}(\mathcal{U}, \mathbf{p}, \mathbf{Q}(t), \widehat{\mathbf{H}}(t))$ and $S^{\mathrm{npr}}_{\max}(t) = \max_{k \in \mathcal{K}_{\mathrm{npr}}} S^{\mathrm{npr}}(\{k\}, \mathbf{Q}(t))$ and chooses to serve the best subset of predictable users if $S^{\mathrm{npr}}_{\max}(t) \leq S^{\mathrm{pr}}_{\max}(t)$, or the best non-predictable user if $S^{\mathrm{npr}}_{\max}(t) > S^{\mathrm{pr}}_{\max}(t)$.

## V. RESULTS AND DISCUSSION

In this section we illustrate the performance advantages of the proposed MU-MIMO scheduling policies over a conventional "mismatched" PFS scheme that treats the existing CSI as if it was perfect. The mismatched scheme computes the ZFBF vectors $\{\mathbf{v}_k\}$ from the CSI matrix $\widehat{\mathbf{H}}(t)$ as described in Section II, and selects the active user subset by maximizing the *mismatched* weighted sum rate $\sum_{k=1}^{K} r_k(t)/\mathcal{T}_k(t)$, where $r_k(t) = \log\left(1 + \frac{|\widehat{\mathbf{h}}_k(t)^{\mathsf{H}}\mathbf{v}_k|^2 p_k}{N_0}\right)$ and where the powers $\{p_k\}$ are computed by waterfilling [4]. The coefficients $\mathcal{T}_k(t)$ represent time-averaged rates, that are updated according to the rule [5], [6]

$$\mathcal{T}_k(t+1) = (1 - 1/t_c)\mathcal{T}_k(t) + (1/t_c)R_k(t)$$

where $R_k(t)$ denotes the actual service rate of user $k$, under the outage or optimistic rate assumption as defined in (4) and in (5). It is well-known that this algorithm approximately maximizes $\sum_k \log \bar{R}_k$ in the case of perfect CSI (i.e., when $\widehat{\mathbf{H}}(t) = \mathbf{H}(t)$ for all $t$), when $t_c$ is very large.

In Section V-A, we consider an idealized setting where all channels are i.i.d. Rayleigh fading, and where a subset $\mathcal{K}_{\mathrm{pr}}$ of users have perfect CSI while the complement set $\mathcal{K}_{\mathrm{npr}}$ has channels completely unknown to the BS. Then, in Section V-B we considered the SCM channel model used in 3GPP standardization [14], and actual channel estimation and prediction schemes. Interestingly, even in this very realistic setting, similar performance trends are observed.

## A. Rayleigh fading

We consider a BS with $M = 4$ antennas and $K = 8$ users, with $\mathcal{K}_{\mathrm{npr}} = \{1, 2\}$. In this section we consider the extreme case where the channel of users 1 and 2 is completely unknown to the BS, while the channels of the other users are perfectly known. All channel vectors are i.i.d. across scheduling slots and in the antenna domain, with elements $\sim \mathcal{CN}(0, 1)$ (independent block-fading with spatially white Rayleigh fading). In this case, the mismatched scheme computes the ZFBF beamforming vectors for users $3, \ldots, 8$ without any orthogonality constraint with respect to the channels of users 1 and 2, since the latter are unknown and isotropically distributed.

We start by illustrating the effect of the constants $V$ and $A_{\max}$ on the scheduling performance. Fig. 1 shows the time evolution of the long-term time-average rates achieved by the proposed approximation of the policy $\gamma^\star$ in the HFS case (max-min throughput), for $A_{\max} = 100$ and $V = 100$, $V = 1000$ when SNR is 20dB. In agreement with Theorem 2, by increasing $V$ the time response of the algorithm becomes slower while the achieved utility function value improves. In general, the two parameters $V$ and $A_{\max}$ should be tuned using the bounds of Theorem 2 and are functions of channel statistics, of $K$ and $M$ and of the SNR $P/N_0$.

Next, we examine the ergodic sum rate and sum log-rate achieved by the new algorithms under PFS and HFS, and compare their performance with that of the mismatched PFS scheme. Figs. 2 and 3 show the scheduling algorithms performance versus SNR in dB, for both the optimistic and the outage rate assumption. The gain of the novel algorithms over mismatched PFS is very large, especially under the outage rate assumption. This fact is understood by considering Fig. 4, showing the users "activity fractions", i.e., the fraction of time slots in which a given user is active. The mismatched PFS allocates a very large fraction of slots to the non-predictable users. This is because if some users have poor quality CSI and the scheduler does not take this explicitly into account, then the fairness induced by the PFS utility function forces the system to serve these users very often. Hence, the unpredictable users "drain" a large fraction of the system capacity despite the fact that there might be a large number of users with very good quality CSI. In contrast, the novel schemes treat the non-predictable users separately, and this has a very significant impact not only on the ergodic rates of these users, but also on the whole system sum rate. It is also interesting to notice that, under the proposed scheduling policies, the gap between optimistic rates and outage rates is very small. This indicates that any suitable fast rate adaptation (e.g., based on rateless coding and/or incremental redundancy ARQ) has only a minor impact on the system performance with respect to a much simpler conventional ARQ scheme.

### B. 3GPP channel model and actual channel prediction schemes

We run extensive experiments based on the so-called "Spatial Channel Model" (SCM) [14]. This channel model is not block-fading and the channel coefficients vary continuously over time. Although this model is frequency selective, we considered a frequency-flat version of the channel corresponding to a single subcarrier of an OFDM system, for consistency with the rest of the paper. For a generic user and antenna (indices are omitted), this channel model yields the time-varying channel coefficients in the form

$$h[i] = \sum_{r=1}^{\eta} A_r e^{j2\pi\zeta_r i} \tag{23}$$

where $\eta$ is the number of impinging scattered wavefronts arriving at the receiver ($\eta = 20$ is specified in [14]), $A_r$ are random complex amplitude coefficients, $\zeta_r$ is the Doppler frequency shift corresponding to the $r$-th wavefront, normalized by the signal bandwidth and $i$ ticks at the symbol rate. In turns, the Doppler shifts are given by $\zeta_r = \frac{f_c v}{c} T_s \cos(\theta_r - \theta_v)$, where $f_c$ is the carrier frequency, $v$ is the mobile speed, $c$ denotes light speed, $T_s$ is the symbol interval, $\theta_r$ is the angle of arrival (AoA) of the $r$-th wavefront, and $\theta_v$ is the mobile azimuth direction.

We assume that a set of $M$ orthogonal downlink pilot symbols are sent by the BS every slot of $T$ symbols. Each user estimates and predicts the channel on the next slot using the pilot symbols. After thorough comparisons of various channel estimation and prediction schemes, not reported here for the sake of space limitation, we report here the results for the two most promising schemes in terms of performance versus complexity. The first scheme consists of a block-by-block prediction based on the parametric estimation of the parameters $\{\eta, A_r, \zeta_r\}$ in (23) using ESPRIT applied to blocks of $N \gg 1$ pilot symbols, as described in [33]. The second scheme is a classical Recursive Least-Squares (RLS), approximating a Wiener MMSE predictor for the channel vector sampled at the pilot-insertion rate $1/T$ [34], [35], [36]. In our simulations we considered a system with parameters given in Table I, that corresponds to a single subcarrier of an OFDM system with 256 subcarriers and bandwidth $256 \times 15 \text{KHz} = 3.84 \text{MHz}$. We compared the two prediction methods by considering the four possible different scenarios of: 1) High speed ($v = 75 \text{km/h}$) vs. low speed ($v = 5 \text{km/h}$) mobiles, and 2) well-separated and packed AoAs of the impinging wavefronts. A known limitation of *any* estimator of a linear combination of sinusoids in noise (see [37]) is that the estimation error increases sharply when the separation between some of the frequency components falls below some minimum resolution that depends on the number of pilots $N$. On the other hand, the RLS prediction error degrades as $\max_r |\zeta_r|$ is non-negligible with respect to the pilot insertion rate $1/T$. It follows that there exists a class of channels with both high mobility and clustered AoAs for which all prediction methods essentially fail. This corresponds to the "non-predictable" users said before.

We considered a BS with $M = 4$ antennas and $K = 8$ UTs. We report only the results for one scenario because of space limitation, but the same trend is observed in a variety of cases (see [15]). We consider high-mobility users with ESPRIT parameter estimation/prediction, where users 1 and 2 have clustered AoAs (given in Table III) and users $3, \ldots, 8$ have well-separated AoAs (given in Table II). The simulation results are obtained by keeping the AoAs fixed, and by averaging with respect to the amplitudes of the SCM model. Fig. 5 shows the average sum-rate for the various scheduling algorithms in this case. We notice that the results for these realistic channel models and actual channel estimation and prediction schemes are in agreement with those for the i.i.d. Rayleigh fading case.

## APPENDIX I

### PROOFS

**Proof of Theorem 1.** First, we show that any arrival rate for which the system is strongly stable must be in $\mathcal{R}$. Suppose that for some uniformly bounded i.i.d. process $\mathbf{A}(t)$ with rate $\boldsymbol{\lambda}$, there exists a policy that stabilizes the system. Using the queue buffer evolution equation (8), assuming $\mathbf{Q}(0) = 0$ for simplicity, and summing with respect to $\tau = 0, \ldots, t-1$ we obtain:

$$\mathbf{Q}(t) \geq \sum_{\tau=0}^{t-1} \mathbf{A}(\tau) - \sum_{\tau=0}^{t-1} \mathbf{R}(\tau) \tag{24}$$

where $\mathbf{R}(t)$ denotes the service rate achieved by the policy. Dividing by $t$, taking expectations and rearranging terms we arrive at:

$$\frac{1}{t} \sum_{\tau=0}^{t-1} \mathbb{E}[\mathbf{A}(\tau)] \leq \frac{1}{t} \sum_{\tau=0}^{t-1} \mathbb{E}[\mathbf{R}(\tau)] + \frac{\mathbb{E}[\mathbf{Q}(t)]}{t} \tag{25}$$

Using the fact that $\mathbb{E}[\mathbf{A}(\tau)] = \boldsymbol{\lambda}$ for all $\tau$, we see that the left hand side of the above bound is equal to $\boldsymbol{\lambda}$. Since strong stability with a finite $A_{max}$ implies mean-rate stability [16], it follows that $\mathbb{E}[\mathbf{Q}(t)/t] \to 0$, and so the final term in (25) converges to the zero vector. Finally, we have $\mathbb{E}[\mathbf{R}(\tau)] \in \mathcal{R}$ for all $\tau$, and hence $\frac{1}{t} \sum_{\tau=0}^{t-1} \mathbb{E}[\mathbf{R}(\tau)] \in \mathcal{R}$ for all $t$ (as this is a convex combination of the vectors $\mathbb{E}[\mathbf{R}(\tau)]$, and $\mathcal{R}$ is a convex set). It follows that $\boldsymbol{\lambda}$ is arbitrarily close to a point in $\mathcal{R}$. Because $\mathcal{R}$ is closed, we conclude that $\boldsymbol{\lambda} \in \mathcal{R}$.

Then, in order to show that $\gamma^*$ stabilizes the system for any $\boldsymbol{\lambda}$ in the interior of $\mathcal{R}$, we will use the Lyapunov drift approach. Let $\mathcal{L}(\mathbf{Q}) = \frac{1}{2} \sum_{k=1}^{K} Q_k^2$ denote a Lyapunov function defined on $\mathbb{R}_+^K$. The corresponding one-step Lyapunov drift is given by

$$\Delta(\mathbf{Q}(t)) = \mathbb{E}\left[\mathcal{L}(\mathbf{Q}(t+1)) - \mathcal{L}(\mathbf{Q}(t))\,|\,\mathbf{Q}(t)\right] \tag{26}$$

The following result is standard (see [16] and references therein):

*Fact 1:* If there exists constants $C > 0$ and $\epsilon > 0$ such that

$$\Delta(\mathbf{Q}(t)) \leq C - \epsilon \sum_{k=1}^{K} Q_k(t) \tag{27}$$

then

$$\limsup_{t\to\infty} \frac{1}{t} \sum_{\tau=0}^{t-1} \sum_{k=1}^{K} \mathbb{E}[Q_k(\tau)] \leq \frac{C}{\epsilon}$$

and hence each queue $Q_k(t)$ is strongly stable. $\diamond$

In order to show that (27) holds in our case, we use (8) and write

$$
\begin{aligned}
Q_k(t+1)^2 &\leq [Q_k(t) - R_k(t)]^2 + A_k^2(t) + 2A_k(t)\max\{0, Q_k(t) - R_k(t)\} \\
&\leq Q_k(t)^2 + R_k(t)^2 + A_k(t)^2 - 2Q_k(t)[R_k(t) - A_k(t)]
\end{aligned}
\tag{28}
$$

Summing with respect to $k$ and applying conditional expectation $\mathbb{E}[\cdot|\mathbf{Q}(t)]$ we arrive at

$$\Delta(\mathbf{Q}(t)) \leq \frac{1}{2}\sum_{k=1}^{K}\mathbb{E}[R_k(t)^2 + A_k(t)^2|\mathbf{Q}(t)] - \sum_{k=1}^{K}Q_k(t)\mathbb{E}[R_k(t) - A_k(t)|\mathbf{Q}(t)] \tag{29}$$

Observing that $R_k(t) \leq \log(1 + |\mathbf{h}_k(t)|^2 P/N_0)$, where the latter is the maximum achievable instantaneous rate for user $k$ under perfect CSI as if it was alone in the system, it follows that

$$\frac{1}{2}\sum_{k=1}^{K}\mathbb{E}[R_k(t)^2 + A_k(t)^2|\mathbf{Q}(t)] \leq \frac{K}{2}\left(A_{\max}^2 + \mathbb{E}[\log^2(1 + |\mathbf{h}_k(t)|^2 P/N_0)]\right) \triangleq C < \infty \tag{30}$$

Next, we shall use the following:

*Lemma 1:* Let the service rates $\{R_k(t)\}$ be obtained by the application of the scheduling policy $\gamma^*$. Then, for any $\bar{\mathbf{R}} \in \mathcal{R}$, we have that

$$\sum_{k=1}^{K}Q_k(t)\mathbb{E}[R_k(t)|\mathbf{Q}(t)] \geq \sum_{k=1}^{K}Q_k(t)\bar{R}_k. \tag{31}$$

*Proof:* Notice that $\mathcal{R}$ is a convex compact region in $\mathbb{R}_+^K$. For any fixed non-negative weight vector $\mathbf{Q}$, the maximum of the linear function $\sum_{k=1}^{K}Q_k r_k$ of $\mathbf{r} \in \mathcal{R}$ is achieved by some $\gamma \in \Gamma(P)$. Hence, for any $\bar{\mathbf{R}} \in \mathcal{R}$ and weight vector $\mathbf{Q}(t)$, there exists $\gamma \in \Gamma(P)$ such that

$$
\begin{aligned}
\sum_{k=1}^{K}Q_k(t)\bar{R}_k &\leq \sum_{k=1}^{K}Q_k(t)\mathbb{E}[R_k(\mathbf{H}(t), \gamma(\widehat{\mathbf{H}}(t)))] \\
&= \sum_{k=1}^{K}Q_k(t)\mathbb{E}\left[\mathbb{E}[R_k(\mathbf{H}(t), \gamma(\widehat{\mathbf{H}}(t)))|\widehat{\mathbf{H}}(t), \gamma]\right] \\
&\leq \mathbb{E}\left[\max_{\boldsymbol{\Sigma}_1,\ldots,\boldsymbol{\Sigma}_K,\mathbf{r}} \sum_{k=1}^{K}Q_k(t)\mathbb{E}\left[R_k(\mathbf{H}(t), \boldsymbol{\Sigma}_1,\ldots,\boldsymbol{\Sigma}_K,\mathbf{r})|\widehat{\mathbf{H}}(t)\right]\Bigg|\mathbf{Q}(t)\right] \\
&= \sum_{k=1}^{K}Q_k(t)\mathbb{E}\left[\mathbb{E}\left[R_k(\mathbf{H}(t), \gamma^*(\widehat{\mathbf{H}}(t)))\Big|\widehat{\mathbf{H}}(t)\right]\Big|\mathbf{Q}(t)\right]
\end{aligned}
\tag{32}
$$

Since we assumed that the service rates $\{R_k(t)\}$ are obtained by applying the policy $\gamma^*$, then, by definition, $\mathbb{E}\left[\mathbb{E}\left[R_k(\mathbf{H}(t), \gamma^*(\widehat{\mathbf{H}}(t)))\Big|\widehat{\mathbf{H}}(t)\right]\Big|\mathbf{Q}(t)\right] = \mathbb{E}\left[R_k(t)|\mathbf{Q}(t)\right]$, and the Lemma is proved. ∎

Now, let $\boldsymbol{\lambda}$ be in the interior of $\mathcal{R}$ and let the service rates $\{R_k(t)\}$ be obtained by $\gamma^*$. Then, there exists a $\epsilon > 0$ such that $\boldsymbol{\lambda} + \epsilon\mathbf{1} \in \mathcal{R}$. Letting $\bar{\mathbf{R}} = \boldsymbol{\lambda} + \epsilon\mathbf{1}$ in (31) and using Lemma 1 we have

$$\sum_{k=1}^{K} Q_k(t)\mathbb{E}\left[R_k(t) - A_k(t)|\mathbf{Q}(t)\right] = \sum_{k=1}^{K} Q_k(t)\left(\mathbb{E}\left[R_k(t)|\mathbf{Q}(t)\right] - \lambda_k\right) \leq \epsilon\sum_{k=1}^{K} Q_k(t) \quad (33)$$

Using (30) and (33) in (29) we find that the condition (27) is satisfied under $\gamma^*$.

**Proof of Theorem 2.** It is convenient to define the quantities $\overline{\mathbf{A}}(t) = \frac{1}{t}\sum_{\tau=0}^{t-1}\mathbb{E}[\mathbf{A}(\tau)]$ and $\overline{\mathbf{R}}(t) = \frac{1}{t}\sum_{\tau=0}^{t-1}\mathbb{E}[\mathbf{R}(\tau)]$, where $\mathbf{A}(t)$ and $\mathbf{R}(t)$ are the virtual arrival process and the service rate vector induced by policy $\gamma^*$. We start with a preliminary fact, the proof of which uses the general bound (25) and the fact that strong stability and uniformly bounded arrival processes implies mean-rate stability (i.e., $\mathbb{E}[\mathbf{Q}(t)]/t \to \mathbf{0}$) [16].

*Fact 2:* Suppose queues $\mathbf{Q}(t)$ are strongly stable and there is a finite upper bound $A_{\max}$ on arrivals every slot. If $g(\cdot)$ is a continuous and entry-wise non-decreasing function, then:

$$\liminf_{t\to\infty} g(\overline{\mathbf{A}}(t)) \leq \liminf_{t\to\infty} g(\overline{\mathbf{R}}(t)) \quad (34)$$

$$\limsup_{t\to\infty} g(\overline{\mathbf{A}}(t)) \leq g(\bar{\mathbf{R}}^\star(A_{\max})) \quad (35)$$

$\diamondsuit$

From (29), (30) and Lemma 1, we can write

$$\Delta(\mathbf{Q}(t)) \leq C - \sum_{k=1}^{K} Q_k(t)\bar{R}_k + \sum_{k=1}^{K} Q_k(t)\mathbb{E}[A_k(t)|\mathbf{Q}(t)] \quad (36)$$

where $\Delta(\mathbf{Q}(t))$ is the Lyapunov drift defined in (26), $C$ is the constant given in (30) and $\bar{\mathbf{R}} = (\bar{R}_1, \ldots, \bar{R}_K)$ is any vector in $\mathcal{R}$. Following the technique of [16], [17], we subtract a term related to the utility function from both sides of (36) to yield:

$$\Delta(\mathbf{Q}(t)) - V\mathbb{E}[g(\mathbf{A}(t))|\mathbf{Q}(t)] \leq C - \sum_{k=1}^{K} Q_k(t)\bar{R}_k + \mathbb{E}\left[\sum_{k=1}^{K} Q_k(t)A_k(t) - Vg(\mathbf{A}(t))\Big|\mathbf{Q}(t)\right]$$

Note from the definition of $\gamma^\star$ that $\mathbf{A}(t)$ is chosen for every $t$ to minimize the right hand side over all vectors $\mathbf{a}$ that satisfy $0 \leq a_k \leq A_{\max}$ for all $k$. Let $\mathbf{z}$ be any vector in $\mathcal{R}$ that satisfies

$0 \leq z_k \leq A_{\max}$ for all $k$. Thus:

$$\Delta(\mathbf{Q}(t)) - V\mathbb{E}[g(\mathbf{A}(t))|\mathbf{Q}(t)] \quad \leq \quad C - \sum_{k=1}^{K} Q_k(t)\bar{R}_k + \sum_{k=1}^{K} Q_k(t)z_k - Vg(\mathbf{z})$$

Taking expectations of both sides of the above inequality and using the law of iterated expectations yields:

$$\mathbb{E}[\mathcal{L}(\mathbf{Q}(t+1))] - \mathbb{E}[\mathcal{L}(\mathbf{Q}(t))] - V\mathbb{E}[g(\mathbf{A}(t))] \quad \leq \quad C - \sum_{k=1}^{K} \mathbb{E}[Q_k(t)](\bar{R}_k - z_k) - Vg(\mathbf{z})$$

For simplicity, assume that $\mathbf{Q}(0) = 0$. The above inequality holds for all $t$. Summing the above over $\tau \in \{0, \ldots, t-1\}$, dividing by $t$, rearranging terms, and using non-negativity of $\mathcal{L}(\cdot)$ gives:

$$\frac{1}{t} \sum_{\tau=0}^{t-1} \sum_{k=1}^{K} \mathbb{E}[Q_k(\tau)](\bar{R}_k - z_k) \leq C + Vg(\overline{\mathbf{A}}(t)) - Vg(\mathbf{z}) \tag{37}$$

where we have used Jensen's inequality in the concave function $g(\cdot)$. The above holds for all $t$, all $\bar{\mathbf{R}} \in \mathcal{R}$, and all $\mathbf{z} \in \mathcal{R}$ such that $0 \leq z_k \leq A_{\max}$ for all $k$. Parts (a) and (b) of Theorem 2 are proven by plugging different values into (37). We first prove part (b).

*Proof of part (b).* Take any point $\mathbf{x} \in \mathcal{R}$ such that $0 \leq x_k \leq A_{\max}$ for all $k$. Choose $\bar{\mathbf{R}} = \mathbf{x}$ and $\mathbf{z} = \beta\mathbf{x}$, for any $\beta \in [0, 1]$. Then from (37) we have:

$$\frac{1}{t} \sum_{\tau=0}^{t-1} \sum_{k=1}^{K} x_k \mathbb{E}[Q_k(\tau)] \leq \frac{C + Vg(\overline{\mathbf{A}}(t)) - Vg(\beta\mathbf{x})}{1 - \beta} \tag{38}$$

At this point, we first prove that the queues are strongly stable and then, using Fact 2, we obtain part (b). Notice that $g(\overline{\mathbf{A}}(t)) \leq g(\mathbf{A}_{\max})$, where $\mathbf{A}_{\max}$ is a vector with all entries equal to $A_{\max}$. Using this bound in (38) and taking a $\limsup$ yields:

$$\limsup_{t \to \infty} \frac{1}{t} \sum_{\tau=0}^{t-1} \sum_{k=1}^{K} x_k \mathbb{E}[Q_k(\tau)] \leq \frac{C + Vg(\mathbf{A}_{\max}) - Vg(\beta\mathbf{x})}{1 - \beta} \tag{39}$$

By assumption, there exists at least one point $\mathbf{r} \in \mathcal{R}$ that has all positive entries and such that $g(\mathbf{r}/2) > -\infty$. Choosing $\beta = 1/2$ and $\mathbf{x} = \mathbf{r}$, it follows that the right-end side of (39) is finite and hence all queues are strongly stable.

Because of strong stability and since the arrival processes are uniformly bounded by $A_{\max} < \infty$ by construction, we can apply inequality (35) of Fact 2 to the right-end side of (38) and obtain the result of part (b).

*Proof of part (a).* We plug $\bar{\mathbf{R}} = \mathbf{z} = \bar{\mathbf{R}}^\star(A_{\max})$ into (37) and obtain:

$$g(\overline{\mathbf{A}}(t)) \geq g(\bar{\mathbf{R}}^\star(A_{\max})) - C/V$$

Taking $\liminf$ and using (34) in Fact 2 yields the result of part (a).

## References

[1] W. Yu and T. Lan, "Transmitter optimization for the multi-antenna downlink with per-antenna power constraints," *IEEE Transactions on Acoustics, Speech, and Signal Processing*, vol. 55, no. 6, pp. 2646 − 2660, June 2007.

[2] A. Wiesel, Y. C. Eldar, and S. Shamai, "Zero-forcing precoding and generalized inverses," *IEEE Transactions on Signal Processing*, vol. 56, no. 9, pp. 4409 − 4418, Sep 2008.

[3] F. Boccardi, F. Tosato, and G. Caire, "Precoding Schemes for the MIMO-GBC," in *Int. Zurich Seminar on Communications*, February 2006, pp. 10 − 13.

[4] G. Dimic and N. Sidiropoulos, "On Downlink Beamforming with Greedy User Selection: Performance Analysis and Simple New Algorithm," *IEEE Trans. on Sig. Proc.*, vol. 53, no. 10, pp. 3857–3868, October 2005.

[5] P.Viswanath, D.N.C.Tse, and R.Laroia, "Opportunistic Beamforming Using Dumb Antennas," *IEEE Trans. on Inform. Theory*, vol. 48, no. 6, June 2002.

[6] V. K. N. Lau, "Proportional Fair SpaceTime Scheduling for Wireless Communications," *IEEE Transactions on Communications*, vol. 53, no. 8, pp. 1353–1360, August 2005.

[7] M.Sharif and B.Hassibi, "On the Capacity of MIMO Broadcast Channel with Partial Side Information," *IEEE Trans. on Inform. Theory*, vol. 51, no. 2, pp. 506 − 522, February 2005.

[8] J. Jose, A. Ashikhmin, P. Whiting, and S. Vishwanath, "Scheduling and precoding in multi-user multiple antenna time division duplex systems," *submitted to IEEE Trans. on Commun., Arxiv preprint arXiv:0812.0621*, 2008.

[9] J. Jose, A. Ashikhmin, T. L. Marzetta, and S. Vishwanath, "Pilot contamination problem in multi-cell tdd systems," *Arxiv preprint arXiv:0901.1703*, 2009.

[10] G. Caire, N. Jindal, M. Kobayashi, and N. Ravindran, "Multiuser MIMO Downlink Made Practical: Achievable Rates with Simple Channel State Estimation and Feedback Schemes," *Submitted to IEEE Trans. Information Theory*, Nov. 2007, Arxiv preprint cs.IT/0711.2642v1.

[11] H. Shirani-Mehr and G. Caire, "Channel State Feedback Schemes for Multiuser MIMO-OFDM Downlink," *Accepted for publication in IEEE Transactions on Communications* , 2009.

[12] H. Bang, T. Ekman, and D. Gesbert, "Channel predictive proportional fair scheduling," *IEEE Transactions on Wireless Communications*, vol. 7, no. 2, pp. 482 − 487, February 2008.

[13] M. Kobayashi, G. Caire, and D. Gesbert, "Transmit diversity versus opportunistic beamforming in data packet mobile downlink transmission," *IEEE Trans. on Commun.*, vol. 55, no. 1, pp. 151 − 157, January 2007.

[14] 3GPP, "Spatial channel model for multiple input multiple output (mimo) simulations," TR 25.996, 2003.

[15] H. Shirani-Mehr, "Mimo downlink with non-perfect channel state information: Prediction, channel state feedback and scheduling," Ph.D. dissertation, in preparation.

[16] L. Georgiadis, M. Neely, and L. Tassiulas, *Resource Allocation and Cross-Layer Control in Wireless Networks*, ser. Foundations and Trends in Networking.   Hanover, MA, USA: Now Publishers Inc., 2006, vol. 1, no. 1.

[17] M. J. Neely, E. Modiano, and C. Li, "Fairness and optimal stochastic control for heterogeneous networks," *IEEE INFOCOM Proceedings*, March 2005.

[18] M. J. Neely, E. Modiano, and C. E. Rohrs, "Dynamic power allocation and routing for time varying wireless networks," *IEEE Journal on Selected Areas in Communications, Special Issue on Wireless Ad-Hoc Networks*, vol. 23, no. 1, pp. 89 − 103, Jan 2005.

[19] J. Zhang, R. Heath Jr., M. Kountouris, and J. G. Andrews, "Mode switching for mimo broadcast channel based on delay and channel quantization," *Submitted to IEEE Transactions on Wireless Communications*, December 2008.

[20] H. Weingarten, Y. Steinberg, and S. Shamai, "The capacity region of the Gaussian multiple-input multiple-output broadcast channel," *IEEE Trans. on Inform. Theory*, vol. 52, no. 9, pp. 3936 – 3964, September 2006.

[21] B. Hochwald, C. Peel, and A. Swindlehurst, "A Vector-Perturbation Technique for Near-Capacity Multiantenna Multiuser Communication-Part II: Perturbation," *IEEE Trans. on Commun.*, vol. 53, no. 3, pp. 537–544, 2005.

[22] G. Caire and S. Shamai, "On the achievable throughput of a multiantenna Gaussian broadcast channel," *IEEE Trans. on Inform. Theory*, vol. 49, no. 7, pp. 1691–1706, 2003.

[23] P. Ding, D. Love, and M. Zoltowski, "Multiple Antenna Broadcast Channels With Shape Feedback and Limited Feedback," *IEEE Trans. on Sig. Proc.*, vol. 55, pp. 3417–3428, 2007.

[24] V. Tarokh, H. Jafarkhani, and A. R. Calderbank, "Space-time block coding for wireless communications: Performance results," *IEEE Journal on Selected Areas in Communications*, vol. 17, no. 3, pp. 451–460, March 1999.

[25] S. M. Alamouti, "A simple transmitter diversity scheme for wireless communications," *IEEE Journal on Selected Areas in Communications*, vol. 16, no. 8, p. 14511458, October 1998.

[26] E. Biglieri, J. Proakis, S. Shamai, and D. di Elettronica, "Fading channels: information-theoretic and communications aspects," *IEEE Trans. on Inform. Theory*, vol. 44, no. 6, pp. 2619–2692, 1998.

[27] P.Bender, P.Black, M.Grob, R.Padovani, N.Sindhushayana, and A.Viterbi, "CDMA/HDR: A bandwidth-efficient high-speed wireless data service for nomadic users," *IEEE Commun. Mag.*, vol. 38, pp. 70–77, July 2000.

[28] H. Holm, G. Oien, M. Alouini, D. Gesbert, and K. Hole, "Optimal design of adaptive coded modulation schemes for maximum average spectral efficiency," in *4th IEEE Workshop on Signal Processing Advances in Wireless Communications, SPAWC 2003*, June 2003, pp. 403 – 407.

[29] H. Kushner and P. Whiting, "Asymptotic properties of proportional-fair sharing algorithms," *Proc. of 40th Annual Allerton Conf. on Communication, Control, and Computing*, 2002.

[30] R. Agrawal and V. Subramanian, "Optimality of certain channel aware scheduling policies," *Proc. 40th Annual Allerton Conference on Communication , Control, and Computing, Monticello, IL*, Oct. 2002.

[31] A. Stolyar, "Greedy primal-dual algorithm for dynamic resource allocation in complex networks," *Queueing Systems*, vol. vol. 54, pp. 203-220, 2006.

[32] A. Eryilmaz and R. Srikant, "Fair resource allocation in wireless networks using queue-length-based scheduling and congestion control," *Proc. IEEE INFOCOM*, March 2005.

[33] I. C. Wong and B. L. Evans, "Sinusoidal Modeling and Adaptive Channel Prediction in Mobile OFDM Systems," *IEEE Transactions on Signal Processing*, vol. 56, no. 41, pp. 1601–1615, April 2008.

[34] I. C. Wong, A. Forenza, R. W. Heath, and B. L. Evans, "Long Range Channel Prediction for Adaptive OFDM Systems," *Proc. IEEE Asilomar Conference on Signals, Systems, and Computers*, vol. 1, pp. 732–736, November 2004.

[35] D. Schafhuber and G. Matz, "Mmse and adaptive prediction of time-varying channels for ofdm systems," *IEEE Transactions on Wireless Communications*, vol. 4, no. 2, p. 593  602, March 2005.

[36] S. Haykin, *Adaptive filter theory (2nd ed.)*.  Upper Saddle River, NJ, USA: Prentice-Hall, Inc., 1991.

[37] D. Rife and R. Boorstyn, "Multiple tone parameter estimation from discrete-time observations," *Bell System Technical Journal*, vol. 55, no. 3, pp. 1389–1410, March 1976.

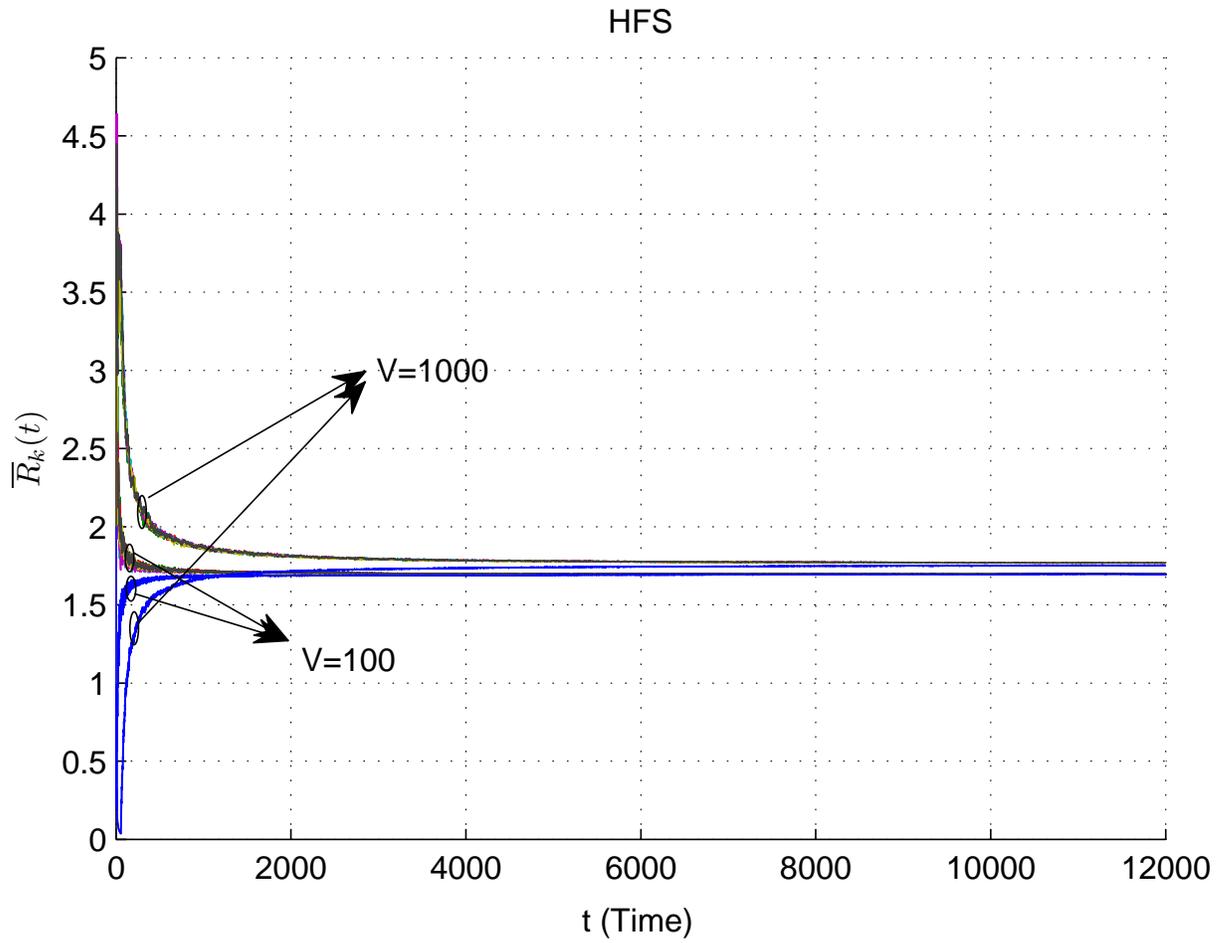

Fig. 1. New HFS, $A_{\max} = 100$, $V = 100$ vs. $V = 1000$.

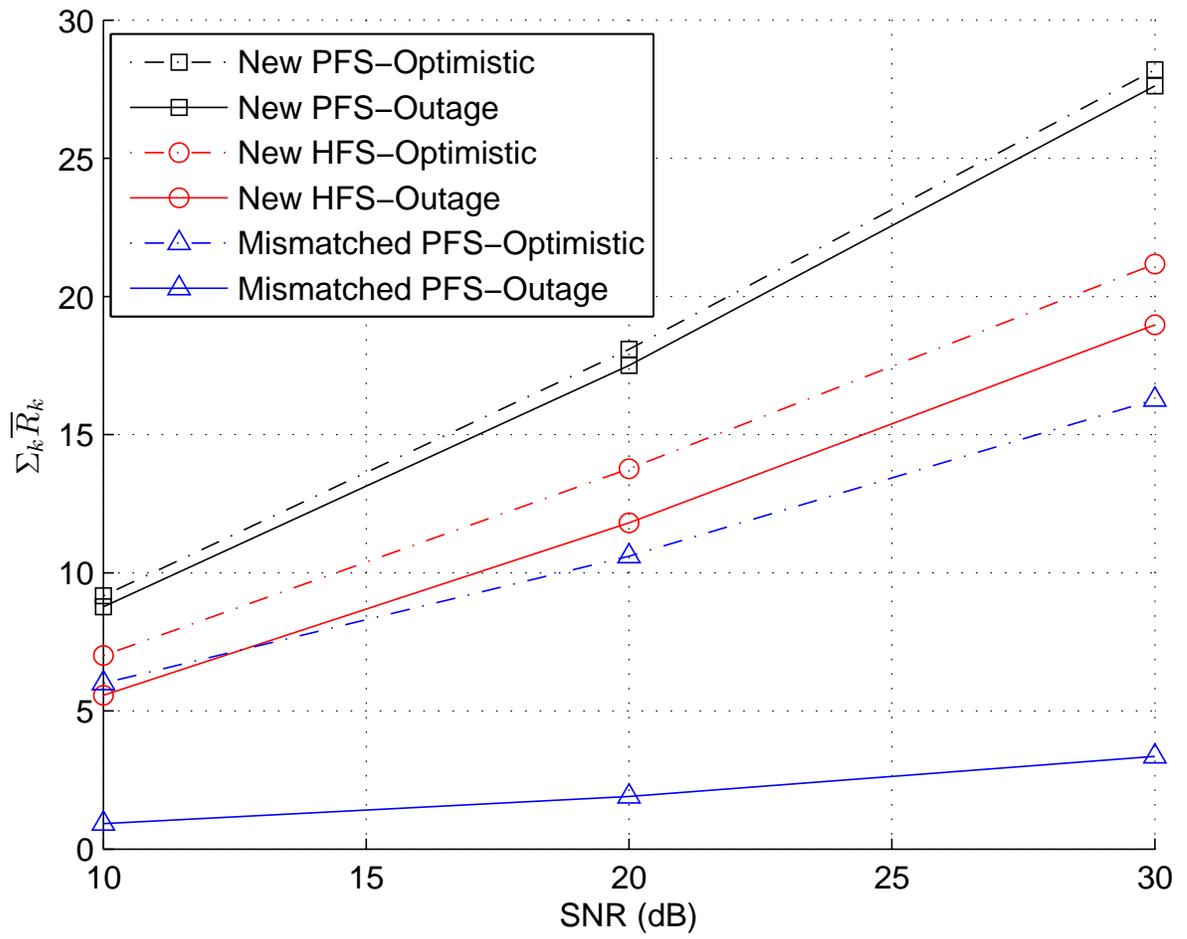

Fig. 2. Ergodic sum rate, Rayleigh fading.

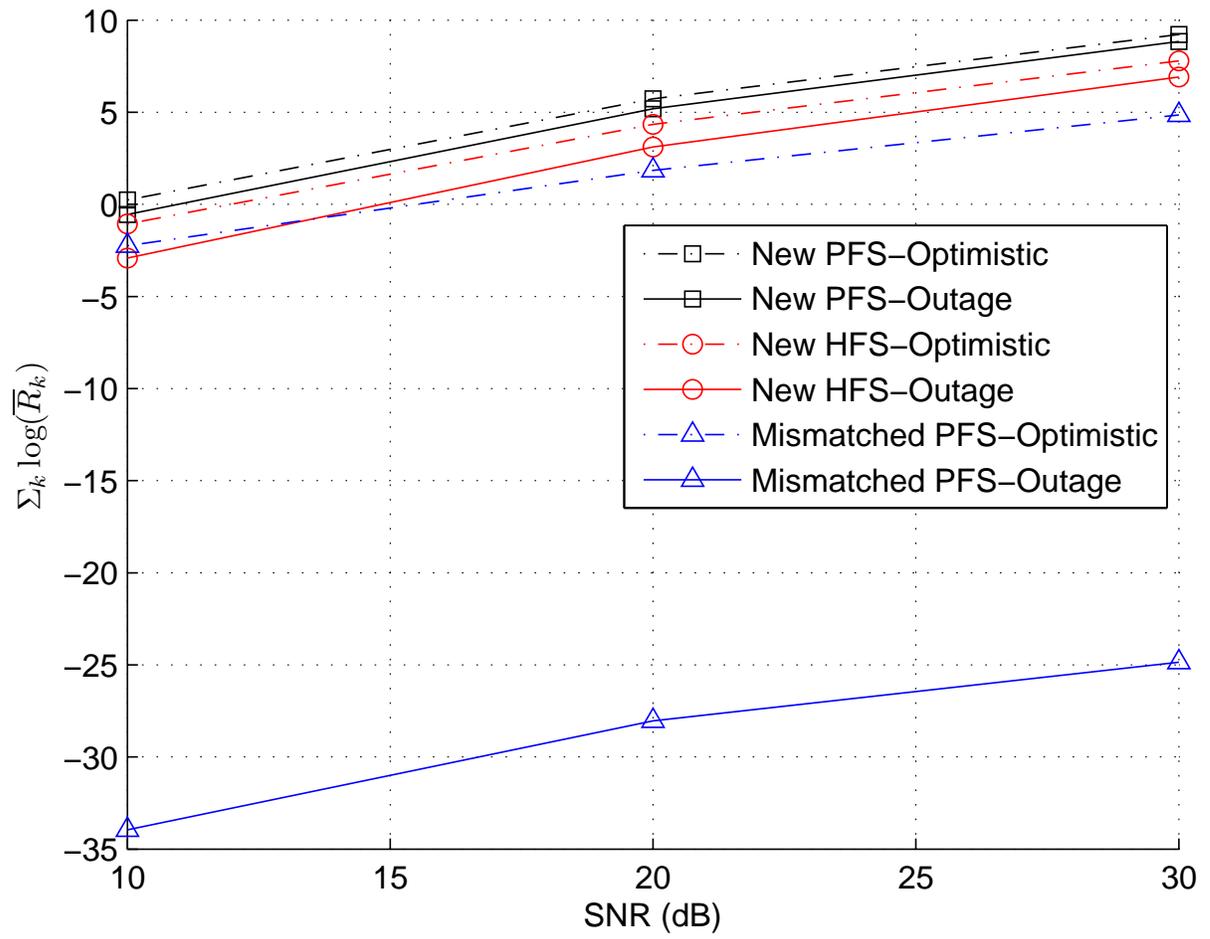

Fig. 3. Sum log ergodic rate, Rayleigh fading.

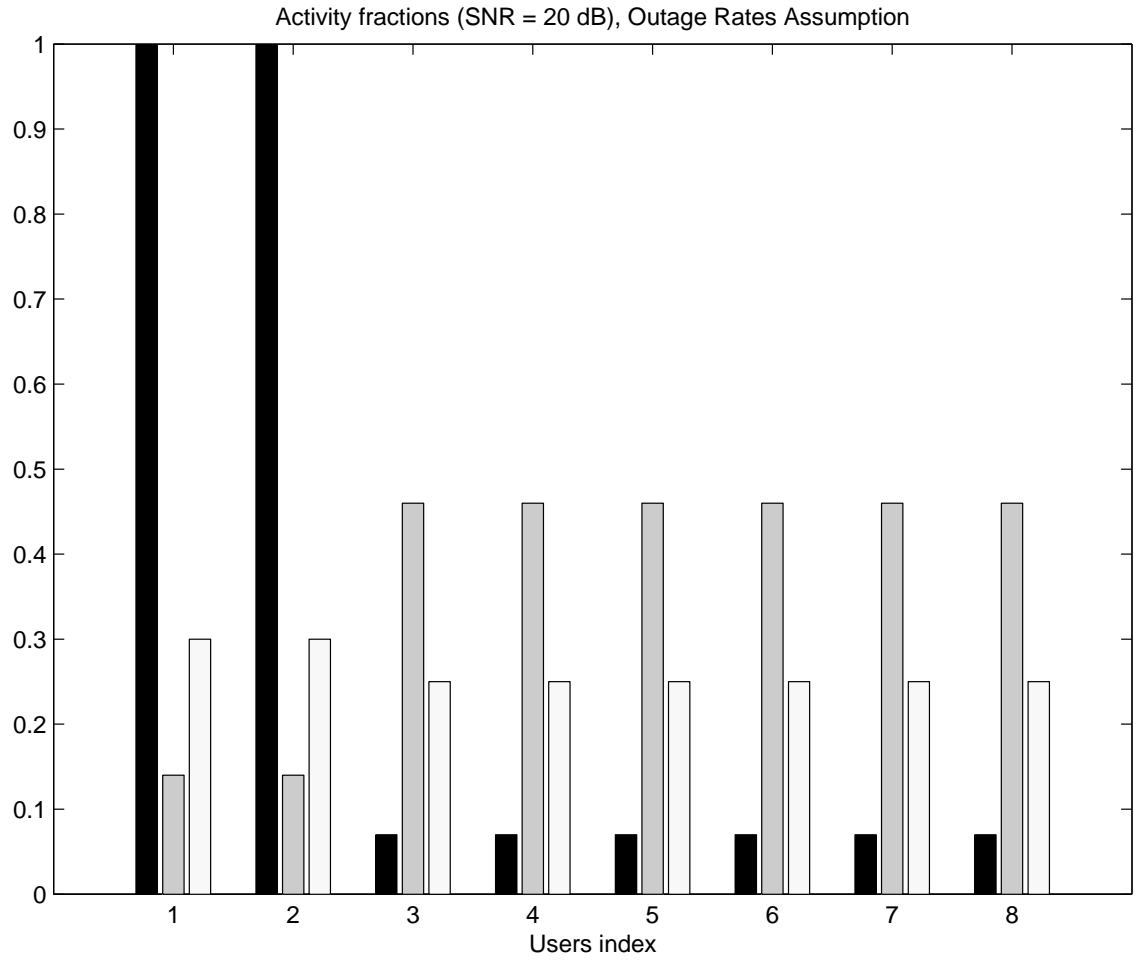

Fig. 4.  Activity fractions at SNR = 20 dB, outage rate assumption, Rayleigh fading (Black: Mismatched PFS, Grey: New PFS; White: New HFS).

TABLE I

Sᴇᴡ Pᴀʀᴀᴍᴇᴛᴇʀꜱ ꜰᴏʀ ꜱɪᴍᴜʟᴀᴛɪᴏɴ

| Description | Value |
|---|---|
| $1/T_s$, symbol rate | 15KHz |
| $f_c$, Carrier frequency | 2.6GHz |
| $N$, Number of pilot symbols | 200 |
| $T$, Pilot symbol spacing | 20 |
| $d_{\min}$, scattering distance | 600m |

TABLE II

Aɴɢʟᴇꜱ ᴏꜰ ᴀʀʀɪᴠᴀʟ ꜰᴏʀ ᴡᴇʟʟ-ꜱᴇᴘᴀʀᴀᴛᴇᴅ ᴄᴀꜱᴇ (ɪɴ ʀᴀᴅɪᴀɴꜱ), $\theta_v = 4.4780$ ʀᴀᴅɪᴀɴꜱ

| $\theta_1$ | $\theta_2$ | $\theta_3$ | $\theta_4$ | $\theta_5$ | $\theta_6$ | $\theta_7$ | $\theta_8$ | $\theta_9$ | $\theta_{10}$ |
|---|---|---|---|---|---|---|---|---|---|
| 4.8328 | 5.2210 | 5.4479 | 5.6090 | 5.7340 | 5.8360 | 5.9223 | 5.9970 | 6.0629 | 6.1219 |
| $\theta_{11}$ | $\theta_{12}$ | $\theta_{13}$ | $\theta_{14}$ | $\theta_{15}$ | $\theta_{16}$ | $\theta_{17}$ | $\theta_{18}$ | $\theta_{19}$ | $\theta_{20}$ |
| 6.1765 | 6.2356 | 6.3015 | 6.3762 | 6.4625 | 6.5644 | 6.6895 | 6.8505 | 7.0774 | 7.4657 |

TABLE III

Angles of arrival for packed case (in radians), $\theta_v = 0.6939$ radians

| $\theta_1$ | $\theta_2$ | $\theta_3$ | $\theta_4$ | $\theta_5$ | $\theta_6$ | $\theta_7$ | $\theta_8$ | $\theta_9$ | $\theta_{10}$ |
|---|---|---|---|---|---|---|---|---|---|
| 3.7263 | 3.6717 | 3.7854 | 3.6127 | 3.8513 | 3.5468 | 3.9260 | 3.4721 | 4.0123 | 3.3858 |
| $\theta_{11}$ | $\theta_{12}$ | $\theta_{13}$ | $\theta_{14}$ | $\theta_{15}$ | $\theta_{16}$ | $\theta_{17}$ | $\theta_{18}$ | $\theta_{19}$ | $\theta_{20}$ |
| 4.1142 | 3.2838 | 4.2393 | 3.1588 | 4.4003 | 2.9977 | 4.6272 | 2.7708 | 5.0155 | 2.3826 |

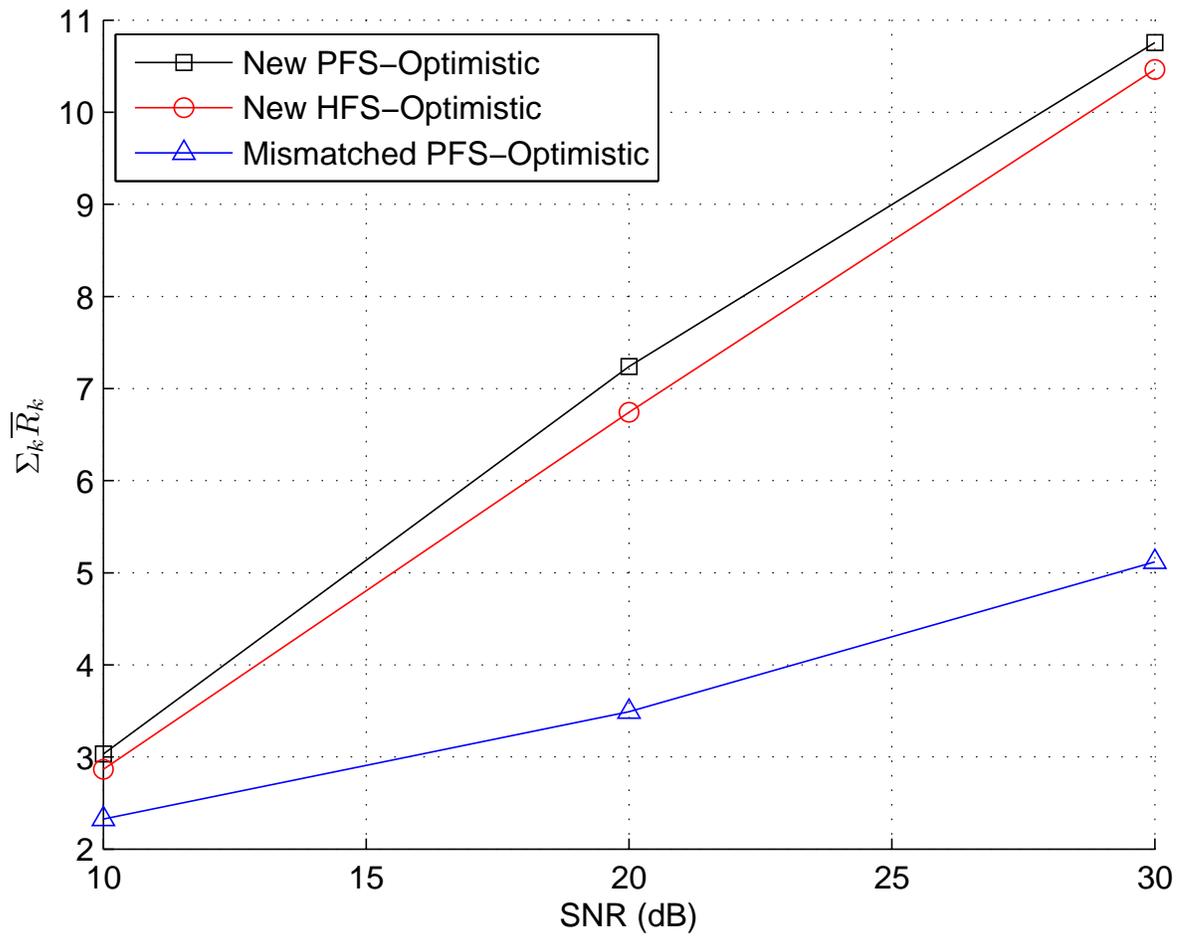

Fig. 5. Average sum rate, SCM channel model, ESPRIT prediction, optimistic rates.